\documentclass[letterpaper,twocolumn,english,aps,prb,groupedaddress,superscriptaddress,showpacs,amsfonts] {revtex4}

\usepackage{epsfig}
\usepackage{graphicx}
\usepackage{dcolumn}
\usepackage{bm}

\usepackage{color}

\DeclareMathAlphabet{\mathpzc}{OT1}{pzc}{m}{it}


\newcommand{\beq}{\begin{equation}}
\newcommand{\eeq}{\end{equation}}
\newcommand{\bea}{\begin{eqnarray}}
\newcommand{\eea}{\end{eqnarray}}

\begin{document}

\title{Thermodynamic Spin Glass Phase Induced by Weak Random Exchange
Disorder in a Classical Spin Liquid: the Case of the Pyrochlore Heisenberg Antiferromagnet}

\author{Ka-Ming Tam}

\affiliation {Department of Physics and Astronomy, University of Waterloo, Waterloo, ON, N2L 3G1, Canada}

\author{Adam J. Hitchcock}

\affiliation {Department of Physics and Astronomy, University of Waterloo, Waterloo, ON, N2L 3G1, Canada}

\author{Michel J. P. Gingras}

\affiliation {Department of Physics and Astronomy, University of Waterloo, Waterloo, ON, N2L 3G1, Canada}

\affiliation {Canadian Institute for Advanced Research, 180 Dundas Street West, Suite 1400, Toronto, ON, M5G 1Z8, Canada}

\date{\today}

\begin{abstract}

The glassy behavior observed in the pyrochlore magnet Y$_{2}$Mo$_{2}$O$_{7}$, 
where the magnetic Mo$^{4+}$ 
ions interact predominantly 
via isotropic nearest neighbor antiferromagnetic exchange, possibly with additional weak 
disorder, is a distinct class of spin glass systems
where frustration is mostly geometrical. A model proposed to describe 
such a spin glass behavior is the Heisenberg model on a pyrochlore lattice 
with random but strictly antiferromagnetic exchange disorder.
In this paper, we provide compelling numerical evidence 
from extensive
Monte Carlo simulations which show that the model exhibits
a finite temperature spin 
glass transition and thus is a realization of a 
 spin glass induced by random weak disorder from spin liquid. 
From our results, we are led to suggest that the spin glass
state of 
 Y$_{2}$Mo$_{2}$O$_{7}$
is driven by effective strong disorder.


\end{abstract}


\maketitle

Most magnetic materials develop long range magnetic order when the temperature is sufficiently low.
There are, however, two prominent exceptions: spin liquid (SL) and spin glass (SG) systems. 
Both of them are commonly found among frustrated 
systems where the ordering tendency is reduced. The SL usually occurs in geometrically frustrated systems, 
whereas the SG arises in random 
frustrated systems due to competing random antiferromagnetic (AFM) and ferromagnetic (FM) couplings~\cite{Binder}. 

The Y$_{2}$Mo$_{2}$O$_{7}$ pyrochlore Heisenberg antiferromagnet~\cite{Y2Mo2O7,Gardner,Gingras-Y2Mo2O7},
 possibly with some form of weak disorder, 
whose origin remains both mysterious and controversial \cite{Booth,Keren,Greedan-PRB}, 
does not fall in the category of conventional SG materials~\cite{Binder}. The Heisenberg AFM model on
 the three-dimensional pyrochlore 
lattice of corner sharing tetrahedra is a classical spin liquid (CSL) with macroscopically degenerate ground states
 which satisfy the 
zero net magnetic
 moment ($\Phi_{\rm t}=0$)
 constraint on each 
tetrahedron \cite{pyro-AFM}. This constraint leads to a gauge theory 
description of the CSL
and the prediction of a power-law decaying spin-spin correlation 
function of ``dipolar'' form \cite{dipolar_form}.
Thus, a clean Heisenberg AFM model on 
the pyrochlore lattice does not display a SG state. 
However, the macroscopic degeneracy 
in this model opens up the interesting 
possibility that weak random disorder in the spin-spin exchange interactions, so small that no
 competing AFM-FM coupling is present in the bare 
Hamiltonian, may be sufficient to induce a SG phase at nonzero temperature.
 The main question that we address in this paper
is whether dense  random weak disorder in the AFM exchange can induce a thermodynamic SG phase
 from a CSL as a case study of what may be 
occurring in Y$_{2}$Mo$_{2}$O$_{7}$~\cite{Y2Mo2O7,Gardner,Gingras-Y2Mo2O7}.

We study the Hamiltonian ${\cal H}$ defined on a pyrochlore lattice
first proposed by
Bellier-Castella {\it et 
al.} \cite{Bellier-Castella}:
\begin{eqnarray}
{\cal H} = {\cal H}_{0} + {\cal H}_{\rm dis}, 
\label{Hamiltonian}
\end{eqnarray}
with
${\cal H}_0 = J_{0}\sum_{\langle i,j \rangle} \bm{S}_{i} \cdot \bm{S}_{j}$ and
${\cal H}_{\rm dis} = \sum_{\langle i,j \rangle} J^{\rm dis}_{ij} \bm{S}_{i} \cdot  \bm{S}_{j}$,
where $\{\bm{S}_{i}\}$ are three component unit vectors and summations are over nearest neighbors. 
${\cal H}_{0}$ is the usual 
pyrochlore Heisenberg AFM model which displays on its own a CSL \cite{pyro-AFM} 
while ${\cal H}_{\rm dis}$ describes the random  disorder which mimics the 
situation in Y$_{2}$Mo$_{2}$O$_{7}$ \cite{Greedan-PRB}. 
Weak disorder here means $|J^{\rm dis}_{ij}| \ll J_0$. 
We set the Boltzmann constant $k_{\rm B}=1$, and also set $J_{0}=1$ 
which serves as overall
 energy scale. The bond disorder is uniformly distributed in the range $J^{\rm  
dis}_{ij}=[-W,W]$ with $W = J_{0}/10$ used in the calculations. 
We refer to ${\cal H}$ in Eq.(1) as the BGHM model~\cite{Bellier-Castella}.

The original work of Bellier-Castella {\it et al.} suggested,
on the basis of measurements of the SG overlap parameter~\cite{Binder}, 
that glassy behavior in ${\cal H}$ sets in at a temperature roughly 
the same as $W$ \cite{Bellier-Castella}. In more recent studies, 
Saunders and Chalker \cite{Saunders} and  Andreanov {\it et al.} \cite{Andreanov} 
computed the SG correlation function and SG susceptibility.
Based on numerical data and analytic arguments, 
the authors of Ref.~[\onlinecite{Andreanov}]
 suggested that there exists a thermodynamic SG transition at a nonzero temperature 
for arbitrary small but nonzero $W$.

SG simulations techniques have significantly improved over the past ten years or 
so~\cite{PT,Ballesteros,Katzgraber,Hasenbusch,Lee1,Lee2,Viet1,Viet2,Fernandez}. 
Recent extensive Monte Carlo simulations employing these improvements
\cite{Lee1,Lee2,Viet1,Viet2,Fernandez}
have led to the 
revision of the old belief that the lower critical dimension, $d_{l}$,
of the Edwards-Anderson (EA) isotropic Heisenberg SG model is above three \cite{Binder}.
There is no rigorous analytic approach to 
determine the $d_{l}$ of the BGHM model \cite{Bellier-Castella,Saunders,Andreanov} 
and the previous MC simulations \cite{Bellier-Castella,Saunders,Andreanov} 
do not come close to the computational standard of recent studies of 
the EA Heisenberg SG \cite{Lee1,Lee2,Viet1,Viet2,Fernandez}. 
Thus, the analytic arguments and 
numerical data
at hand can hardly provide convincing evidence for a thermodynamic SG phase
in ${\cal H}$.
The BGHM model,
with its underlying CSL state in the disorder-free regime, 
as well as its broad relevance 
to the SG behavior observed in numerous 
geometrically frustrated magnetic materials, 
make it a model of fundamental significance in the field
of frustrated magnetism.
It is therefore
important to carefully 
assess whether ${\cal H}$ sustains a thermodynamic SG phase at nonzero 
temperature, and to reach such a conclusion on the basis of numerics 
that approach the standard of SG simulations of EA models \cite{Lee1,Lee2,Viet1,Viet2,Fernandez}. 

We first summarize the details of our MC simulations. A $16$ site 
cubic unit cell for the pyrochlore lattice is used for generating cubic
simulation cells with $N=16L^3$ spins with $L=4$, $6$, $8$. 
One Metropolis sweep, $2L$ over-relaxation sweeps and one parallel tempering 
swap \cite{PT} is defined as one elementary MC step (MCS).
 The temperatures explored for each simulation are 
$T^{(n)} = T_{\rm min}\alpha^{n}$ where 
$T_{\rm min}$ is the lowest temperature considered and $n \in [0,N_{T}-1]$, 
where $N_{T}$ is the number of thermal replicas. Thus the highest 
temperature is $T_{\rm max} = T_{\rm min}\alpha^{N_{T}-1}$ and
 $\alpha = \root (N_{T}-1) \of {T_{\rm max}/T_{\rm min}}$. The error 
bars are sample-to-sample fluctuations calculated via the jackknife method. 
Table \ref{para_table} lists the parameters used in our MC simulations.

\begin{table}[htbp]
\centering
\begin{tabular}{c c c c c c c }
\hline
$L$ & $T_{\rm min}$ & $T_{\rm max}$ & $N_{T}$ & $N_{\rm MCS}^{\rm equil.}$ & $N_{\rm MCS}^{\rm measurement}$   & $N_{\rm samples}$ \\ \hline
$ 4 $& $0.012$ & $0.028$ & $24$ & $2\times10^4$ & $2\times10^4$         & 1750 \\
$ 6 $& $0.012$ & $0.028$ & $30$ & $4\times10^4$ & $2\times10^4$         & 1763 \\
$ 8$ & $0.012$ & $0.028$ & $46$ & $8\times10^4$ & $2\times10^4$       & 1613 \\ \hline
\end{tabular}
\caption{Parameters of the Monte Carlo simulations.}
\label{para_table}
\end{table}

To characterize a putative SG phase, we use a parameter defined as the overlap between two thermal replicas with the same  
realization of random couplings, $\{J^{\rm dis}_{ij}\}$,
\begin{eqnarray}
q^{\mu,\nu}_{\rm SG}(\mathbf{k})\equiv\frac{1}{N}\sum_{i=1}^{N} S^{(1)}_{i,\mu} S^{(2)}_{i,\nu} 
\exp(i\mathbf{k}\cdot\mathbf{r}_{i}),
\end{eqnarray}
where $S^{(1)}_{i,\mu}$ and $S^{(2)}_{i,\mu}$ are the spin components for replicas $(1)$ and $(2)$, respectively.
It has been proposed that there is no SG transition in
isotropic Heisenberg SG systems but that,
instead, the freezing is in the chiral sector \cite{Kawamura1}.
Latest simulations \cite{Lee1,Lee2,Viet1,Viet2,Fernandez}
suggest that both chiral glass (CG) and SG transitions occur at finite temperature,
but there is no consensus whether the CG critical
temperature ($T_{\rm CG}$) is higher or equal to that of the SG ($T_{\rm SG}$).
Since ${\cal H}$ has isotropic Heisenberg spins,
there is no  obvious reason to exclude the possibility of a CG transition.
To monitor CG correlations, we consider two chirality parameters.
The first one is defined along bonds, 
\begin{eqnarray}  
q_{\rm CG1}(\mathbf{k})\equiv\frac{1}{3N}\sum_{i=1}^{N} 
\,
\sum_{\hat{\delta}=\hat{\epsilon}_{i,1}; \hat{\epsilon}_{i,2}; \hat{\epsilon}_{i,3}} 
\!\!\!\!\!\!
\kappa^{(1)}_{1;i,\hat{\delta}}\kappa^{(2)}_{1;i,\hat{\delta}}
\exp(i\mathbf{k}\cdot\mathbf{r}_{i}), 
\end{eqnarray}
where $\kappa_{1;i,\hat{\delta}} = \bm{S}_{(i,{\hat{\delta}})} \cdot (\bm{S}_{i} \times \bm{S}_{(i,-\hat{\delta})})$. 
The $\hat{\epsilon}_{i,1}$, $\hat{\epsilon}_{i,2}$ and $\hat{\epsilon}_{i,3}$ 
are vectors pointing from site $i$, to its three nearest neighbor 
sites in the same tetrahedron; $(i,\hat{\delta}), i, (i,-\hat{\delta})$ are the indices for the three sites lying along the 
direction $\hat{\delta}$ (see Appendix).
The second one is defined on the triangular faces of individual tetrahedra,
\begin{eqnarray}
q_{\rm CG2}(\mathbf{k})\equiv\frac{1}{2N}\sum_{\omega=1}^{2N} {\kappa}^{(1)}_{2;\omega}{\kappa}^{(2)}_{2;\omega}
\exp(i\mathbf{k}\cdot \tilde{\mathbf{r}}_{\omega}),
\end{eqnarray}
where ${\kappa}_{2;\omega} = \bm{S}_{\omega,a} \cdot (\bm{S}_{\omega,b} \times \bm{S}_{\omega,c})$; 
$(\omega,a),(\omega,b),(\omega,c)$ 
are the indices for the three sites of the triangular face $\omega$ and 
$\tilde{\mathbf{r}}_{\omega} = (\tilde{\mathbf{r}}_{\omega,a} 
+ \tilde{\mathbf{r}}_{\omega,b} 
+ \tilde{\mathbf{r}}_{\omega,c})/3$. 
The corresponding susceptibilities are obtained from the order parameters, $\chi_{\rm SG}(\mathbf{k}) = 
N\sum_{\mu,\nu=x,y,z}[\langle|q^{\mu,\nu}_{\rm SG}(\mathbf{k})|^{2} \rangle]$; 
$\chi_{\rm CG1}(\mathbf{k}) = 3N [\langle|q_{\rm 
CG1}(\mathbf{k})|^{2} \rangle] / \overline{\chi_{1}}^{4}$, 
$\chi_{\rm CG2}(\mathbf{k}) = 2N [\langle|q_{\rm CG2}(\mathbf{k})|^{2} \rangle] / 
\overline{\chi_{2}}^{4}$, where $\langle...\rangle$ and $[...]$ 
denote the thermal average and disorder average, respectively. 
As the local chirality variables are not fixed 
to be unity as is
 the case of the spin variables \cite{Viet1,Viet2}, the CG1 susceptibility is normalized by 
$\overline{\chi_{1}}^{4} = (\overline{\chi_{1}}^{2})^{2}$, where 
$\overline{\chi_{1}}^{2} \equiv \frac{1}{3N} \sum_{i=1}^{N} 
\sum_{\hat{\delta}=\hat{\epsilon}_{i,1},\hat{\epsilon}_{i,2},\hat{\epsilon}_{i,3}}  
[\langle{\kappa}_{1;i,\hat{\delta}}^{2}\rangle]$; 
and, similarly, the CG2 susceptibility is normalized by 
$\overline{\chi_{2}}^{4} = (\overline{\chi_{2}}^{2})^{2}$, where 
$ \overline{\chi_{2}}^{2} \equiv \frac{1}{2N} \sum_{\omega=1}^{2N}   
[\langle{\kappa}_{2;\omega}^{2}\rangle]$ (see Appendix).

Assuming that the susceptibilities follow an Ornstein-Zernike form \cite{Ballesteros}, 
the correlation lengths $\xi_{\rm SG}$ and $\xi_{\rm CG1,CG2}$
can be determined via 
 $ \xi_{g}(L) = \frac{1}{2 {\rm sin} (|\mathbf{k}|/2)} 
 \left( \frac{\chi_{g}(\mathbf{0})}{\chi_{g}(\mathbf{k})}-1 \right) 
^{1/2}$, where $g \equiv$ SG, CG1 or CG2 and
 $\mathbf{k}\equiv 2\pi\hat{x}/L$ is  one of the 
smallest wave vectors for system size $L$   \cite{Lee1,Lee2,Viet1,Viet2,Fernandez}.
 The $\xi_g$'s 
divided by $L$ 
 should be scale invariant at their respective critical point.
 The crossing of $\xi_g/L$ is therefore 
a sensitive criterion to test for a glass transition. 
The correlation lengths and susceptibilities should finite-size scale as
 $ \xi_{g}(L)/L = X[(T-T_{g})L^{1/\nu_{g}}] \label{Xi_scaling}$ and 
$\chi_{g}(T,L)L^{\eta_{g}-2} = Y[(T-T_{g})L^{1/\nu_{g}}] \label{Chi_scaling}$, respectively.
To check that thermodynamic equilibrium was reached, we verified that
$\xi_g$ becomes independent of simulation time for the largest system size
and lowest temperature considered.

\begin{figure}[bth]
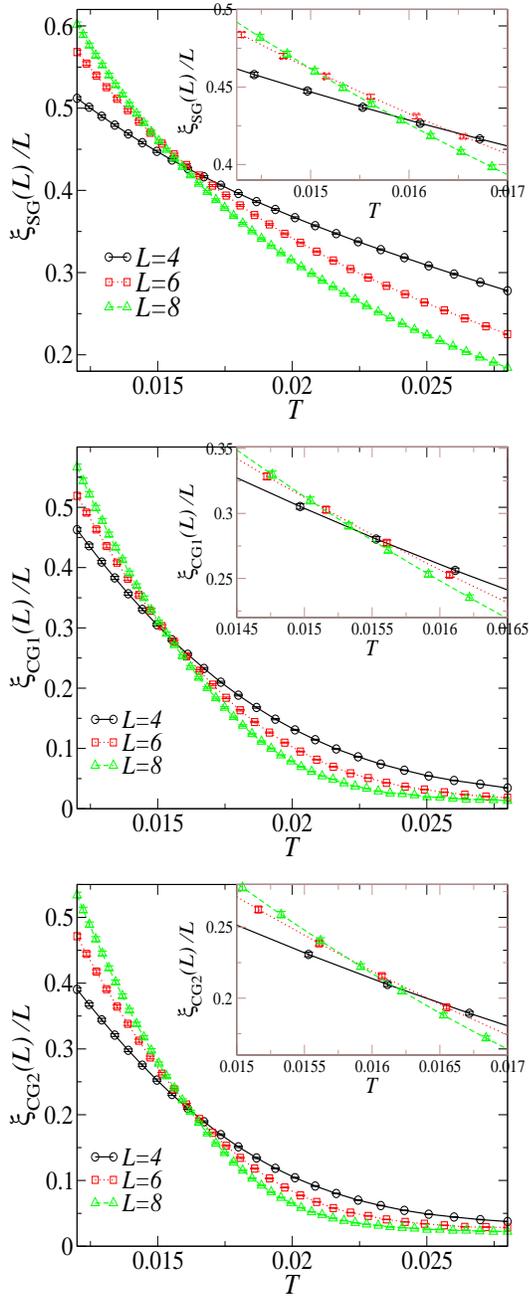

\includegraphics*[height=0.245\textheight,width=0.415\textwidth, viewport=0 0 750 515,clip]{corr_sg.eps}
\includegraphics*[height=0.245\textheight,width=0.415\textwidth, viewport=0 0 750 515,clip]{corr_cg1.eps}
\includegraphics*[height=0.245\textheight,width=0.415\textwidth, viewport=0 0 750 515,clip]{corr_cg2.eps}
\caption{(Color online). Correlation lengths for spin glass (top) and chiral glass (middle and bottom). 
The insets show the  details close to the crossing points.}
\label{correlation_lengths}
\end{figure}

We first present the SG and CG correlation lengths in Fig. \ref{correlation_lengths}. 
It is fairly clear from these results that the SG and CG correlation lengths 
for different system sizes tend to cross 
in a narrow range of temperatures compatible with a
nonzero critical temperature for both SG and CG ($T_g \approx 0.016$). 

\begin{figure}[bth]
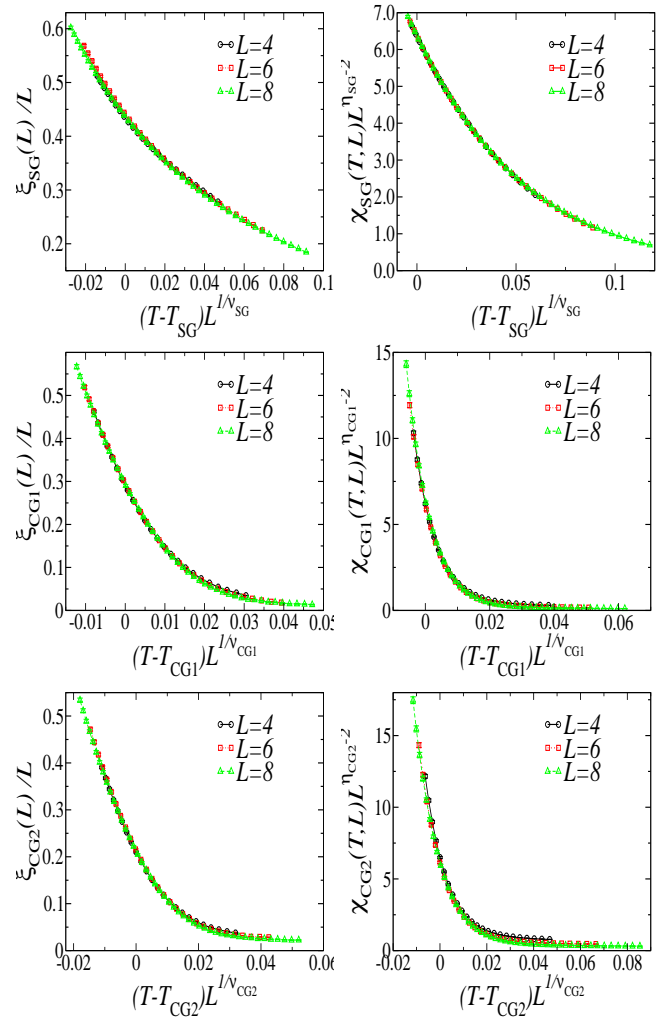

\includegraphics*[height=0.19\textheight,width=0.235\textwidth, viewport=5 0 720 560,clip]{scaled_corr_sgs.eps} 
\includegraphics*[height=0.19\textheight,width=0.235\textwidth, viewport=5 0 720 560,clip]{scaled_chi_sgs.eps}
\includegraphics*[height=0.19\textheight,width=0.235\textwidth, viewport=5 0 720 560,clip]{scaled_corr_cg1s.eps} 
\includegraphics*[height=0.19\textheight,width=0.235\textwidth, viewport=5 0 720 560,clip]{scaled_chi_cg1s.eps}
\includegraphics*[height=0.19\textheight,width=0.235\textwidth, viewport=5 0 720 560,clip]{scaled_corr_cg2s.eps} 
\includegraphics*[height=0.19\textheight,width=0.235\textwidth, viewport=5 0 720 560,clip]{scaled_chi_cg2s.eps}
\caption{(Color online). Left column: Correlation lengths for spin glass (top) 
and chiral glass (middle and bottom) vs $(T-T_{g})L^{1/\nu_{g}}$. 
Critical temperatures and scaling exponents 
are $T_{\rm SG} = 0.0157$, $\nu_{\rm SG} = 1.037$ for SG;  
$T_{\rm CG1} = 0.0153$, $\nu_{\rm CG1} = 1.585$ and $T_{\rm CG2} = 0.0161$, $\nu_{\rm CG2} = 1.408$  for CG, 
obtained by fitting the data to the scaling function $X$. 
Right column: $\chi_{g}(T,L)L^{\eta_{g}-2}$ for spin glass (top) and 
chiral glass (middle and bottom) vs $(T-T_{g})L^{1/\nu_{g}}$. 
Critical temperatures and scaling exponents are
$T_{\rm SG} = 0.0126$, $\nu_{\rm SG} = 1.024$, $\eta_{\rm SG} = -0.292$ for SG; 
$T_{\rm CG1} = 0.0134$, $\nu_{\rm CG1} = 1.439$, $\eta_{\rm CG1} = 0.621$ 
and $T_{\rm CG2} = 0.0139$, $\nu_{\rm CG2} = 1.155$, $\eta_{\rm CG2} = 0.698$ for CG, 
obtained by fitting the data to the scaling function $Y$. 
The numbers listed are the actual values used in these scaling plots. 
Corrections to scaling are expected to be sizable, therefore no error
 bar is provided, while the exponents lie within the
range estimated for the 3D EA Heisenberg model. }

\label{scaled}
\end{figure}

We then employ a scaling scheme which assumes that the correlation 
lengths finite-size scale as the 
scaling function $X$ given above and fit the data in the temperature 
range ($0.012 \leq T \leq 0.020$) by parametrizing $X$ as
polynomials $X(z)=\sum_{m=0,1,...,5}c_{m}(z-z_{0})^{m}$. 
The merit function $\Delta$,
$\Delta \equiv \sum_{\rm MC \: data}[X(z)/(\xi_{g,L}/L)-1]^{2}$, 
is minimized numerically to obtain the coefficients 
$c_{m}$, $z_{0}$, critical temperature $T_{g}$ and  exponent $\nu_{g}$. 
Figure \ref{scaled} shows $\xi_{g}(L)/L$ versus the 
scaling parameter $z \equiv (T-T_{g})L^{1/\nu_{g}}$ ($T_{g}$ and $\nu_{g}$ 
are listed in the caption of Fig. \ref{scaled}.)
The scaling exponents determined both for the SG and CG are far from 
those of the 3D EA Ising model obtained from correlation length 
scaling ($\nu_{\rm SG} \approx 2.44$) \cite{Katzgraber,Hasenbusch}, but roughly comparable
(within $\sim 20$\%) to those of the 3D EA Heisenberg model
 \cite{Viet1,Viet2,Lee1,Lee2,Fernandez,auxmaterial}.

\begin{figure}[bth]
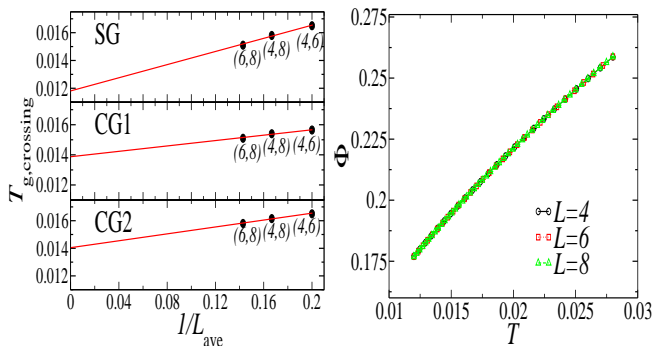

\includegraphics*[height=0.19\textheight,width=0.235\textwidth, viewport=0 4 760 524,clip]{Tg_scale.eps}
\includegraphics*[height=0.19\textheight,width=0.235\textwidth, viewport=0 0 740 510,clip]{flux.eps}
\caption{(Color online). 
Left: Correlation length crossing temperatures 
for SG, CG1 and CG2 for different pairs of system sizes $(L_{1},L_{2})$. 
The  vertical axes are the crossing temperatures and the horizontal axis is the inverse 
of the average system sizes given by $1/L_{\rm ave} = 2 / (L_{1}+L_{2})$. 
 The lines are guides for linear extrapolations to the limit $L_{\rm ave}
\to \infty$, that is assuming 
$T_{g}^{*}(L_{1},L_{2}) - T_{g} \propto L_{\rm ave}^{-\theta}$ with $\theta=1$.
Right: The average tetrahedra moment, $\Phi$, vs temperature, $T$.
}
\label{Tg_scale-Flux}
\end{figure}

In the above scaling analyses, we assumed that there is a common crossing point for all
system sizes. Realistically, 
the critical temperature obtained this way for fair system sizes ($N\leq 8192$) 
should represent {\it an upper bound} for the true critical temperature in the thermodynamic limit.
For example, in the latest simulations of the 3D EA Heisenberg model, 
it was found that 
 scaling corrections are large and
 that the $\xi_g$ crossings are pushed to lower, albeit 
non-zero, temperatures as the system size increases \cite{Lee1,Lee2,Fernandez,Viet1,Viet2}. 
We show in Fig. \ref{Tg_scale-Flux} 
the evolution of the correlation 
lengths crossing temperatures $T_{g}^{*}$ for different pairs of 
system sizes 
$(L_{1},L_{2})$ as 
a function of the inverse of their average size given by $1/L_{\rm ave} = 2 / (L_{1}+L_{2})$.
These show that $0 < T_{\rm SG} \lesssim T_{\rm CG1} \approx T_{\rm CG2}$ as $L_{\rm ave}
\rightarrow \infty$.

We use the same procedure as for the correlation length data collapse 
to fit the susceptibilities into the scaling function, $Y$, in order to 
determine $T_{g}$, $\nu_{g}$ and $\eta_{g}$. Figure \ref{scaled} 
shows $\chi_{g}(T,L) L^{\eta_{g}-2}$ versus the scaling parameter 
$z=(T-T_{g})L^{1/\nu_{g}}$. (The $T_{g}$, $\nu_{g}$, and $\eta_{g}$ 
are listed in the caption of Fig. \ref{scaled}.) 
The $\eta_{g}$ and $\nu_{g}$ values are again fairly comparable 
with those obtained in the latest studies of the 
3D EA Heisenberg model \cite{Lee1,Lee2,Viet1,Viet2,Fernandez,auxmaterial}, 
providing evidence for a common SG universality class for the BGHM model 
and the 3D EA Heisenberg model.

A recent study aimed at describing the SG in ${\cal H}$ assumes that the power-law 
correlation of the CSL \cite{dipolar_form}
 is maintained despite
the random disorder and thus the spins can be thought of as interacting via an effective 
projected interaction matrix of a long range ``dipolar'' 
form \cite{Andreanov,Saunders} 
as a consequence of the zero net magnetic moment ($\Phi_{\rm t}$=0) 
condition on each tetrahedron. To investigate this description,
we calculate the average tetrahedra moment,
$\Phi \equiv \left[ \left\langle \frac{\sum_{t} |\sum_{i=1,...,4} 
\bm{S}_{(t,i)}|}  {N_{\rm tetrahedron}} \right\rangle \right]$, 
where the outer sum is over all tetrahedra, the inner sum is over the four spins 
in each tetrahedron, and $N_{\rm tetrahedron}=N/2$ is the total number of tetrahedra.
 We show $\Phi$ as a function of temperature in Fig. 
\ref{Tg_scale-Flux}. First, we find that it changes very little with system size,
 as it is not a critical quantity. Second, it decreases with decreasing temperature. 
Most importantly, near the  crossing temperatures ($T_{g}\approx 0.016$), $\Phi$ is finite and 
of the order of $W/J_{0}$ which implies the existence of ``defect'' tetrahedra 
with $\Phi_{\rm t} \neq 0$ and, 
consequently, the destruction of infinite-range power-law correlations \cite{dipolar_form}.
This is further supported by the reasonably good data collapse for the $\xi_g$ correlation lengths extracted
from an Ornstein-Zernike form, which would likely not be correct if there
were a CSL phase with extended power-law correlations 
intervening between the paramagnetic phase and the SG phase.

For Y$_{2}$Mo$_{2}$O$_{7}$, the experimentally determined Curie-Weiss and SG temperatures
are 
$\Theta_{\rm CW}^{\rm exp.} \approx -200$ K 
\cite{Y2Mo2O7,Gardner} and 
$T_{\rm SG}^{\rm exp.} \approx 22.5$  K \cite{Gingras-Y2Mo2O7}, respectively. 
The nearest neighbor coordination number on the pyrochlore lattice is $z=6$ and the spin $S$ 
of magnetic Mo$^{4+}$ is $S=1$. 
Therefore, with
$J_{0} \sim J_{0}^{\rm mic.} S^2$ and
$\Theta_{\rm CW}^{\rm exp.} = J_{0}^{\rm mic.} z S(S+1)/3 \approx 200{\rm K}$, we get
$J_{0}^{\rm mic.}\approx 50$ K 
and the ratio $T_{\rm SG}^{\rm exp.}/(J_{0}^{\rm mic.}S^2) \approx 0.45$. 
This is much higher than that obtained for 
the BGHM model for which we found 
above $T_{\rm SG}/J_{0} \sim 0.01 - 0.02$ for $W/J_{0} = 0.1$. 
This  suggests that the glass transition of
Y$_{2}$Mo$_{2}$O$_{7}$ is not due to 
weak random disorder as in ${\cal H}$,  but rather to
{\it very strong effective} disorder.
One plausible scenario is that perturbations beyond nearest-neighbor Heisenberg exchange $J_0$ 
disrupt the perfect degeneracy of the CSL phase and induce short range AFM order above $T_g$,
as observed in a neutron scattering study \cite{Gardner}. 
If the growth of AFM order is forestalled
due to some form of random disorder \cite{Booth,Keren,Greedan-PRB}, 
the SG behavior of Y$_{2}$Mo$_{2}$O$_{7}$ should likely be described in terms of 
a ``cluster-glass" model \cite{Binder}.

In conclusion, our MC simulations of the 
BGHM model of Eq. (1) provide compelling evidence for 
a thermodynamic SG phase induced by weak random 
disorder in a classical spin liquid of a highly frustrated system. 
From our work, it appears very likely that the SG behavior in Y$_{2}$Mo$_{2}$O$_{7}$ is
not due to weak and dense random disorder but rather via an effective strong disorder 
whose microscopic 
origin requires further investigation.

This work was funded by the NSERC of Canada, 
the Canada Research Chair Program (M.G., Tier 1) and SHARCNET. 
We thank H. Kawamura for 
encouraging us in studying this problem and for his useful 
comments on our manuscript, and
J. Chalker, P. Holdsworth, P. McClarty and P. Stasiak for useful discussions. 


\appendix

\section{Spin and Chiral Glass Susceptibilities}
\label{sec:general}

This appendix discusses the details of the definitions of the spin glass (SG) 
and two different chiral glass (CG) overlap parameters -- 
CG1 defined along the bonds and CG2 defined on the triangular faces of the
tetrahedra that from a pyrochlore lattice. 
In addition, a table for critical exponents 
obtained from previous Monte Carlo
studies of the three-dimensional (3D) Edwards-Anderson (EA) Heisenberg SG
model is provided for comparison with the critical exponents 
obtained for the BGHM model studied in this paper.

The spin glass (SG) overlap is defined as the overlap between two thermal replicas with the same  
realization of random couplings $\{J^{\rm dis}_{ij}\}$,
\begin{eqnarray}
q^{\mu,\nu}_{\rm SG}(\mathbf{k})\equiv\frac{1}{N}\sum_{i=1}^{N} S^{(1)}_{i,\mu} S^{(2)}_{i,\nu} 
\exp(i\mathbf{k}\cdot\mathbf{r}_{i}),
\end{eqnarray}
where $S^{(1)}_{i,\mu}$ and $S^{(2)}_{i,\mu}$ are the spin components of the two replicas. For Ising spins, this is 
the usual parameter used to monitor the spin freezing. The situation is, however, more complicated for Heisenberg spins
\cite{Kawamura1}.

It has been proposed that there is no SG transition in isotropic Heisenberg SG systems but, 
instead, that the freezing is in the chiral sector 
\cite{Kawamura1}. There have been many investigations on the spin-chirality coupling/decoupling
 scenario for the 3D EA Heisenberg model
\cite{Viet1,Viet2,Fernandez,KawamuraXY,Kawamura-Tanemura-1991,Kawamura1,Kawamura_chirality,Kawamura2,Kawamura3,Kawamura4,Kawamura-review1,Kawamura-review2,Campbell-review,Petit-2002,Hukushima1,Hukushima2}, but
there is so far no consensus
\cite{Lee1,Lee2,Viet1,Viet2,Fernandez,Kawamura-review1}. 
While recent studies tend to agree that both the 
critical temperature for SG ($T_{\rm SG}$) and chiral glass ($T_{\rm CG}$) 
are non-zero \cite{Lee1,Lee2,Viet1,Viet2,Fernandez}, whether $T_{\rm CG}=T_{SG}$ 
\cite{Lee1,Fernandez} or $T_{\rm CG} > T_{\rm SG}$ \cite{Viet1,Viet2} is still under active debate. 
Since the model we study possesses spins with isotropic 
Heisenberg exchange, there is no a priori reason to rule out the possibility of a chiral glass (CG) transition. 

\begin{figure}[bth]
\includegraphics*[height=0.16\textheight,width=0.29\textwidth, viewport=150 50 750 500,clip]{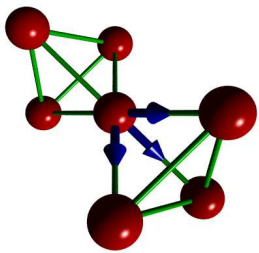}
\caption{(Color online).
 The figure illustrates the vectors for defining the chirality at one of the sites.
 The blue color arrows show the direction 
of the vectors $\hat{e}_{i,1}$, $\hat{e}_{i,2}$ and $\hat{e}_{i,3}$ for the site $i$ (in the middle) shared by two tetrahedra.}
\label{tetrahedron}
\end{figure}

To monitor the CG we probe two different chirality variables.
 The first one is defined along the bonds. This is a generalization of the definition 
of chirality variables for the 3D EA Heisenberg model on a simple cubic lattice.
 Similarly to the simple cubic lattice, three are three axes passing 
through each site. As three spins are needed to define the chirality, 
the natural choice is to pick a spin and its two nearest neighbors 
spins along one of the axes to define a chirality variables. 
As there are three axes passing thorough each site, therefore there are three 
chirality variables for each site. We denote this definition of chirality overlap as CG1. 
\begin{eqnarray}  
q_{\rm CG1}(\mathbf{k})\equiv\frac{1}{3N}\sum_{i=1}^{N} \sum_{\hat{\delta}=\hat{\epsilon}_{i,1},\hat{\epsilon}_{i,2}, 
\hat{\epsilon}_{i,3}} 
\kappa^{(1)}_{1;i,\hat{\delta}}\kappa^{(2)}_{1;i,\hat{\delta}}\exp(i\mathbf{k}\cdot\mathbf{r}_{i}), 
\end{eqnarray}
where $\kappa_{1;i,\hat{\delta}} = \bm{S}_{(i,{\hat{\delta}})} \cdot (\bm{S}_{i} \times \bm{S}_{(i,-\hat{\delta})})$.
The $\hat{\epsilon}_{i,1}$, $\hat{\epsilon}_{i,2}$ and $\hat{\epsilon}_{i,3}$ are vectors pointing from site $i$, 
to its three nearest neighbor
sites in the same tetrahedron, see Fig. \ref{tetrahedron}. 
$(i,\hat{\delta}), i, (i,-\hat{\delta})$ are the indices for the three sites lying 
along $\hat{\delta}$. The normalization factor $1/3N$ is introduced to account for the $3N$ chirality 
variables for this definition 
because of the three chirality variables at each site. We note that a slightly different definition, 
which treats the CG overlap as a three 
components object has been used in a study of 3D EA Heisenberg model \cite{Fernandez}.

The second chirality overlap parameter
is defined on the triangular faces of the tetrahedra.
The pyrochlore lattice is composed of 
corner-sharing 
tetrahedra, and each tetrahedron has $4$ triangular faces. Therefore, another natural choice 
is to define the chirality variables  on each face of the tetrahedra. 
In a lattice with $N$ sites, there are $N/2$ tetrahedra, and each tetrahedron 
has $4$ faces. Therefore, there are in total 
$2N$ chirality variables within this definition. 
We denote this definition of chirality as CG2, with:
\begin{eqnarray}
q_{\rm CG2}(\mathbf{k})\equiv\frac{1}{2N}\sum_{\omega=1}^{2N} 
{\kappa}^{(1)}_{2;\omega}{\kappa}^{(2)}_{2;\omega}\exp(i\mathbf{k}\cdot 
\tilde{\mathbf{r}}_{\omega}),
\end{eqnarray}
where ${\kappa}_{2;\omega} = \bm{S}_{\omega,a} \cdot (\bm{S}_{\omega,b} \times \bm{S}_{\omega,c})$;
 $(\omega,a),(\omega,b),(\omega,c)$ are the 
indices for the three sites of the triangular face 
$\omega$ and $\tilde{\mathbf{r}}_{\omega} = (\tilde{\mathbf{r}}_{\omega,a} + 
\tilde{\mathbf{r}}_{\omega,b} + \tilde{\mathbf{r}}_{\omega,c})/3$.
 The normalization factor $1/2N$ is introduced to account for the $2N$ chirality 
variables for this definition since there are $2N$ triangular faces.

The corresponding susceptibilities are obtained from the above order parameters, 
\begin{eqnarray}  
\chi_{\rm SG}(\mathbf{k})  & = &  N\sum_{\mu,\nu=x,y,z}[\langle|q^{\mu,\nu}_{\rm SG}(\mathbf{k})|^{2} \rangle]; \\ 
\chi_{\rm CG1}(\mathbf{k}) & = &  3N [\langle|q_{\rm CG1}(\mathbf{k})|^{2} \rangle] / \overline{\chi_{1}}^{4}, \\
\chi_{\rm CG2}(\mathbf{k}) & = &  2N [\langle|q_{\rm CG2}(\mathbf{k})|^{2} \rangle] / \overline{\chi_{2}}^{4},
\end{eqnarray}
where $\langle...\rangle$ and $[...]$ denote the thermal average and disorder average
 respectively. As the local chirality variables are not fixed 
to be unity in contrast of the spin variables \cite{Hukushima2,Viet1,Viet2}, 
the CG1 susceptibility is normalized by 
$\overline{\chi_{1}}^{4} = (\overline{\chi_{1}}^{2})^{2}$, where 
\begin{eqnarray} 
\overline{\chi_{1}}^{2} \equiv \frac{1}{3N} \sum_{i=1}^{N}
 \sum_{\hat{\delta}=\hat{\epsilon}_{i,1},\hat{\epsilon}_{i,2},\hat{\epsilon}_{i,3}}  
[\langle{\kappa}_{1;i,\hat{\delta}}^{2}\rangle] ,  
\end{eqnarray} 
and the CG2 susceptibility is normalized by 
$\overline{\chi_{2}}^{4} = (\overline{\chi_{2}}^{2})^{2}$, where 
\begin{eqnarray} 
\overline{\chi_{2}}^{2} \equiv \frac{1}{2N} \sum_{\omega=1}^{2N}   [\langle{\kappa}_{2;\omega}^{2}\rangle].
\end{eqnarray}


The exponents  obtained for the BGHM model
are fairly
comparable with that of the 3D EA Heisenberg spin glass model \cite{Edwards-Anderson}. 
For comparison of our results with the 
3D EA Heisenberg model, we compile a selection of critical temperatures
and critical exponents in Table \ref{3DEA_table}. 
For the 3D EA Ising model, see Table I in Ref.~[\onlinecite{Katzgraber}].

\newpage

\begin{widetext}
\begin{table*}[t]

\begin{tabular}{|c |c |c |c |c |c |c |c |}
\hline
reference & randomness type & $T_{\rm SG}$   & $\nu_{\rm SG}$  & $\eta_{\rm SG}$ &  $T_{\rm CG}$   & $\nu_{\rm CG}$    & $\eta_{\rm CG}$ \\ 
\hline

Kawamura\cite{Kawamura4} & Gaussian        & NA         & NA          & NA          & $0.157 \pm 0.01$ & NA    & NA \\ 
(1998)&& & & & & & \\ \hline

Hukushima and Kawamura\cite{Hukushima1} & Gaussian & NA & NA & NA & $0.160\pm0.005$ & 1.2 & 0.8 \\ 
(2000)&& & & & & & \\ \hline

Endoh, {\it et al.}\cite{Endoh}  & Bimodal & $0.19\pm0.02$ & NA & NA & NA & NA & NA \\ 
(2001)&& & & & & & \\ \hline

Matsubara, {\it et al.} \cite{Matsubara1} & Bimodal & 0.18 & NA&NA & NA& NA& NA\\
(2001)&& & & & & & \\ \hline

Nakamura and Endoh\cite{Nakamura1}&Bimodal & $0.21^{+0.01}_{-0.03}$ & $1.1\pm0.2$ & $0.27^{\#}$ & $0.22^{+0.01}_{-0.04}$ & NA & 
NA  \\ 
(2002)&& & & & & & \\ \hline

Lee and Young\cite{Lee1} &Gaussian & $0.16\pm0.02$ & $1.1\pm0.2$ & NA & $T_{\rm CG}=T_{\rm SG}$ & $1.3\pm0.3$ & NA \\ 
(2003)&& & & & & & \\ \hline

Nakamura, {\it et al.}\cite{Nakamura2} &Bimodal & $0.20\pm0.02$ & $0.8\pm0.2$ & $-0.375^{\#}$ & NA  & NA & NA \\ 
(2003)&& & & & & & \\ \hline

Hukushima and Kawamura\cite{Hukushima2}&Bimodal &NA & NA & NA & $0.19\pm0.01$ & $1.3\pm0.2$ & $0.8\pm0.2$ \\ 
(2005)&& & & & & & \\ \hline

Viet and Kawamura \cite{Viet2} &Gaussian & $0.125^{+0.006}_{-0.012}$ & NA & 
$\lesssim -0.30$& $0.143\pm0.003$ & $1.4\pm0.2$ & $0.6\pm0.2$ \\ 
(2009)&& & & & & & \\ \hline

Fernandez, {\it et al.} \cite{Fernandez}&Gaussian & $0.120^{+0.010}_{-0.004}$ & $1.49\pm0.13^{\ast}$ & $-0.19\pm0.02^{\ast}$ & $T_{\rm CG}=T_{\rm 
SG}$ 
& $1.30\pm0.08^{\ast}$ & $0.56\pm0.04^{\ast}$ \\ 
(2009)&& & & & & & \\ \hline

This work ($\xi/L$ scaling) & pyrochlore & $0.0157$ & $1.037$ & NA & $0.0153$(CG1) & $1.585$(CG1)& NA \\ 
& $\pm 0.1J_{0}$ uniform & & & &$0.0161$(CG2)& $1.408$(CG2)& \\ \hline

This work ($\chi$ scaling) & pyrochlore & $0.0126$ & $1.024$ & $-0.292$ & $0.0134$(CG1) & $1.439$(CG1)& $0.621$(CG1) \\ 
& $\pm 0.1J_{0}$ uniform & & & &$0.0139$(CG2) &$1.155$(CG2) & $0.698$(CG2)\\ \hline

\end{tabular}

\caption{Selection of critical temperatures and exponents of the three-dimensional Edwards-Anderson Heisenberg 
model for the simple cubic lattice. 
For 
the Edwards-Anderson Ising model see Table I in Ref.~[\onlinecite{Katzgraber}].
 NA is a shorthand for not available. The last two rows are the estimates 
from this work. 
The variance of the random coupling distribution 
is $1$ for all the studies on the 
three-dimensional Edwards-Anderson Heisenberg 
model listed in this table, and the variance of the model
 we study is $1/3\times10^{-2}$. 
(The variance is defined as $\int (x-\overline{x})^{2} P(x) dx$, where $P(x)$ is the 
distribution function 
and $\overline{x} = \int x P(x) 
dx$.) The entries with ``$\#$" are not quoted in the original paper but are estimated 
via the relation 
$\gamma=(2-\eta)\nu$. The 
entries with ``$\ast$" are from the quotient method \cite{Binder-FSS,Ballesteros-FSS} for 
scaling analysis of $L=24$
 and $L=48$ systems on a simple 
cubic 
lattice.}

\label{3DEA_table}

\end{table*}

\end{widetext}

\vfill{ }


\end{document}